# Yelp Dataset Analysis using Scalable Big Data

Mohsen Alam, Benjamin Cevallos, Oscar Flores, Randall Lunetto, Kotaro Yayoshi, Jongwook Woo
Department of Information Systems,
California State University Los Angeles
{malam, bcevall, oflores4, rlunett , kyayosh, jwoo5}@calstatela.edu

**Abstract:** Yelp has served and will continue to serve as a data-driven application. Yelp has published a dataset containing business information, reviews, user information, and check-in information. This paper will examine this dataset to provide descriptive analytics to understand business performance, geo-spatial distribution of businesses, reviewers' rating and other characteristics, and temporal distribution of check-ins in business premises. With these analysis we are able to establish that yelp reviews, tips, elite users and check ins have started to plummet over the years. Coincidentally, the paper also establishes that Canadians have a more stable star ratings as well as sentiment ratings when compared to Americans.

## 1. Introduction

Yelp is a common database used by the public that allows the users to view specific data on business. Through the Yelp database you can pull statistical data that can serve us in the future data analysis. In this project we will utilize geo-spatial and temporal distributions to provide data analysis on businesses. This project provides information about tip sentiment analysis which will allow the users to see just how well a business is. Secondly, we have decided to analyze the count per year which gives us a clear understanding of what user, and how many counts of that type of user ratings occur over a span of years. Ratings category count also provides us with information about what the highest rated region is. Lastly, we can also utilize pulling specific data such as what places are the top-rated places for children. With the given different variations of these descriptive analysis, geo-spatial analysis, and temporal analysis, we can use data that is analyzed to determine how well a business is doing as well as anticipate how well it might do in the future.

## 2. Related Work

Geo-Temporal analysis is a process that allows to characterize subsets of items in a geographic database, for example events, with respect to the similarity of their location and timestamp. It is information derived from an analysis of images and data associated with a particular location. It uses imagery to survey and assess human activity and physical geography anywhere on Earth (Figure 1).

We recreated the geo-temporal data presentation, which Ruchi et al did [1]. The methods and data types were similar. As well as the utilization of 3D maps feature to graphically represent the data. However, we present Yelp feature performance and Star Ratings.

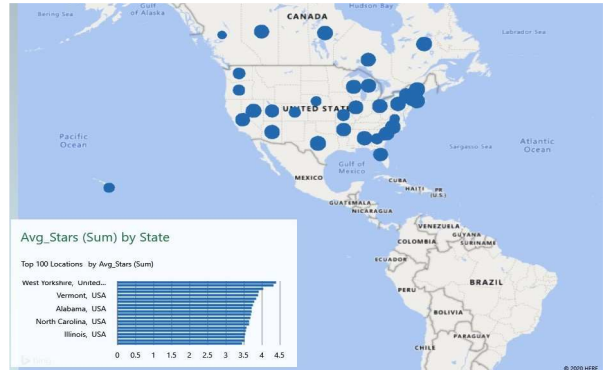
Figure 1. Average Yelp stars per state

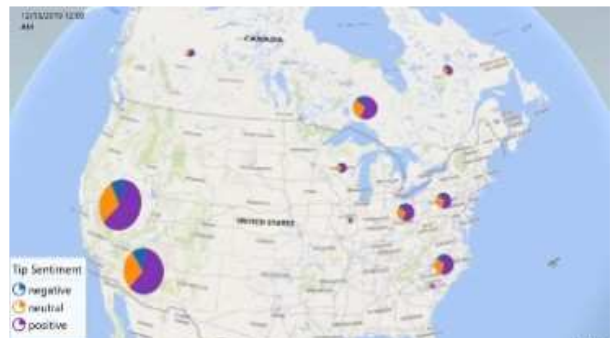
Figure 2. Tip Sentiment of the United States and Canada

For our presentations, we utilized the Big Data techniques for data analysis and graphically display geo-temporal data sets.

Looking at the wide range and diversity of the types of data presented. We can surmise that this analysis method proved to be very robust and could be utilized in a myriad of ways to manipulate data sets in order to graphically visualize conclusions based on certain data outputs. Although context is still required to give meaning to the conclusions shown by the data. The graphical nature allows people to better process the information and form correlations about the data.

## 3. Background

Sentiment analysis, also known as opinion mining, has been a popular research method in the natural language processing field. The goal of sentiment analysis is to define automatic tools to be able to extract subjective information such as opinions and sentiments, from text in natural language. In order to generate data that can be utilized by decision support systems.

Sentiment analysis is often used by companies to quantify general social media opinion. One of the simplest and most common sentiment analysis methods is to classify words as

This study was supported by Oracle Cloud Innovation Accelerator. Oracle Big Data Cloud Service was used in the data analysis.

"positive" or "negative", then to average the values of each word to categorize the entire document.

Griffo [2] has worked on Twitter sentiment analysis in Hive, and we based our work on his project.

His method was to create a counter and name it as polarity then increment or decrement based on the number of positive or negative values there were.

When a 'negative' was triggered, it would decrement the polarity counter by 1 while when there was a 'positive' trigger it would increment the counter by 1.

After which, there would be a small algorithm named tweets_sentiment where the sum of the polarity would be compared with a 0 value.

If the sum of the polarity was greater than zero, it would be labeled as positive. If the sum of the polarity was less than zero, it would be labeled as negative.

For the purposes of this assignment our analysis counts the number of times the words "positive", "negative" or "neutral" are used. The main difference between our codes is that Umberto's utilizes a sentiment dictionary and includes an extra algorithm to delineate between positive and negative words. Ours is a more direct and literal approach for the purposes of displaying the purpose of this type of analysis.

## 4. Specifications

The Yelp dataset provided includes five JSON files which accumulated to the file size of 9.8GB in total. These five JSON files consist of the following: Business.json, Check-in.json, Review.json, User.json, and Tips.json files. The data provided by these json files cover 10 metropolitan areas, 209,393 businesses and 8,021,122 reviews. All of this data spans over fifteen years, ranging from the year 2004 to the year 2019. The total time allotted to create and populate primary tables from these raw JSON files took 161.284 seconds which is approximately 3 minutes total.

We produced three analyses. The first being the Yelp feature performance analysis which the output file consisted of a total output of 16 rows, a file size of 18kb with the total output time of 188.179 seconds which equates to approximately 3 minutes. Secondly, the Yelp tip sentiment output file consisted of a total of 90,950 rows with the data size of 2.48MB and a computation time of 364.896 seconds, summing up to approximately 6 minutes in total. Lastly, the Yelp star-rating analysis output file consisted of a total of 9,832 rows, a file size of 1MB and a computation time of 56.855 seconds. It is to be noted that all these files were converted into xlsx files to be able to convert such big data over into actual usable data.

Table 1 below shows the specifications for our Oracle cluster used.

Table 1. Oracle Cluster Specifications

| Version | 20.3-3-20 |
|---|---|
| Total Nodes | 3 |
| Total Node Memory Size | 180GB |
| Total OCPUs | 12 |
| Total Storage | 957GB |

## 5. Our Work

### 5.1 Data Sources

The dataset used for all of our analyses is provided by Yelp.com [3]

The dictionary file used for our sentiment analysis is provided by CalState LA BigDAI (HiPIC) [4].

The Yelp Dataset [3] uses abbreviations to refer to states, which works fine in Excel 3D Map for known states in the United States and Canada. However, [3] contains some abbreviations for some locations in the United Kingdom and Ireland, which are not recognized by Excel 3D Map. Therefore, we created a mapper text file to increase map quality which translates abbreviations to qualified country and state names [5].

### 5.2 Preparing the Dataset and Primary Tables

We downloaded the Yelp Dataset [3]. We then uploaded the data to the Oracle Big Data Cloud Service (Oracle BDCE) and uncompressed dataset tar file. This resulted in uncompressed files which we uploaded into the Hadoop Distributed File System (HDFS). We then created the primary tables in Hive from the CSV files, which enabled us to utilize Hive Query Language to produce our analyses. The output tables we produced were then copied from HDFS to the local file system which produced output files to be copied into our workstations. Finally, we utilized Excel's 3D maps to generate visualizations of our output tables. The workflow described above is illustrated below (Figure 3).

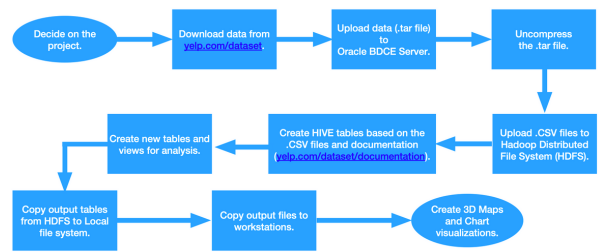

Figure 3. Workflow

### 5.2 Yelp Feature Performance

Our first analysis is on "Yelp Feature Performance". Yelp's platform services millions of users with a variety of features such as, 'Check-Ins', 'Reviews', 'Tips', as well as user status features such as 'New Users' and 'Elite Users'. The goal for our performance analysis is to measure and illustrate how these five features trend over time.

The workflow to produce this analysis is illustrated below in Figure 4.

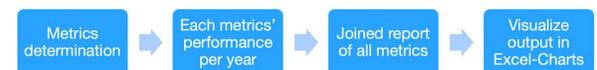

Figure 4. Workflow: Yelp Feature Performance Analysis

We extracted the data from all five of the tables in the dataset: business, check-ins, reviews, tips, and users. We then calculated the count of how many 'New Users', 'Elite

This study was supported by Oracle Cloud Innovation Accelerator. Oracle Big Data Cloud Service was used in the data analysis.

Users', 'Reviews', 'Tips' and 'Check-Ins' to measure performance throughout each year. We visualized this output table via Excel-Charts. The line graph shown in Figure 5 illustrates the trend line of each feature's performance.

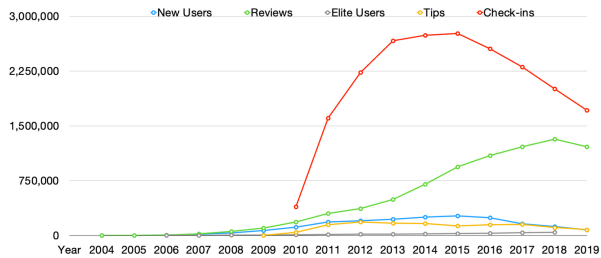
Figure 5. Yelp Feature Performance Line Graph

Key findings are:
i. 'Check-ins' had the highest count and most dramatic drop off;
ii. Four of the five Yelp features measured are trending downwards;
iii. 'Elite Users' is the only feature that is constantly increasing over time but at the slowest rate on average.

The overall output data for Yelp feature performance consists of 16 total rows of data (Table 2). The size of the output file is 18KB and the total computation time in Hive is ~3 minutes (188.179 seconds).

Table 2. Output Table for Yelp Feature Performance

| Year | New Users | Reviews | Elite Users | Tips | Check-Ins |
|---|---|---|---|---|---|
| 2004 | 82 | 12 | \N | \N | \N |
| 2005 | 1,022 | 875 | \N | \N | \N |
| 2006 | 6,052 | 5,030 | 896 | \N | \N |
| 2007 | 17,155 | 21,130 | 2,368 | \N | \N |
| 2008 | 34,327 | 56,996 | 3,592 | \N | \N |
| 2009 | 68,314 | 100,760 | 6,369 | 957 | \N |
| 2010 | 115,106 | 186,752 | 10,238 | 41,922 | 393,953 |
| 2011 | 185,076 | 302,523 | 12,809 | 146,532 | 1,608,736 |
| 2012 | 203,180 | 367,367 | 17,362 | 185,961 | 2,233,001 |
| 2013 | 221,380 | 491,678 | 18,223 | 167,643 | 2,665,596 |
| 2014 | 250,827 | 702,060 | 20,508 | 163,943 | 2,742,368 |
| 2015 | 267,267 | 940,603 | 26,409 | 130,844 | 2,766,769 |
| 2016 | 241,414 | 1,094,154 | 32,128 | 145,569 | 2,560,414 |
| 2017 | 158,881 | 1,217,292 | 38,645 | 151,006 | 2,307,315 |
| 2018 | 122,892 | 1,318,054 | 43,026 | 107,826 | 2,008,051 |
| 2019 | 75,728 | 1,215,836 | \N | 78,558 | 1,717,574 |

## 5.3 Yelp Tip Sentiment
The second analysis is on "Yelp Tip Sentiment". A 'Tip' on Yelp enables its users to leave a short comment or feedback for any business on the Yelp platform [3]. The goal for this analysis is to evaluate the sentiment of user's 'Tips' over time and compare these trends across all geographic locations within the dataset.

The workflow to produce this analysis is illustrated in Figure 6. First, we created a dictionary table based on [4]. Then, we broke down the 'Tips' sentences text into individual words. These words were then assigned a sentiment value of either positive (1), negative (-1) or a neutral (0).

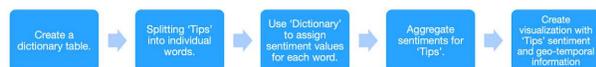
Figure 6 - Workflow: Tip Sentiment Analysis

The sentiments were then aggregated and illustrated as a geo-temporal visualization via Excel's 3D Maps (Figure 7).

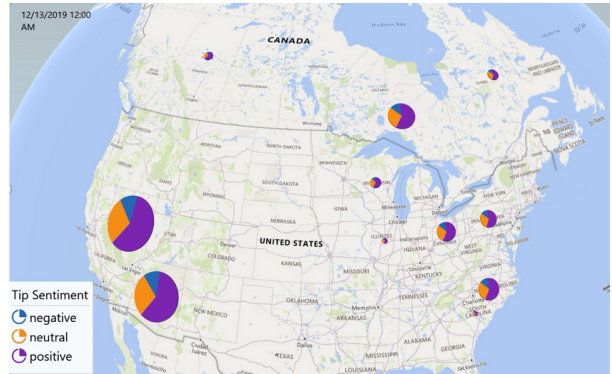
Figure 7 - Geo-Temporal Visualization of Tip Sentiment

This is an animated visualization which allows us to see how user sentiment trends over time across all geographical locations within our dataset.
Key findings are:
i. Overall state-wide sentiment breakdown is: 58.80% Positive, 28.57% Neutral, 12.63% Negative;
ii. Most states overall conform to this trend with low variance (Figure 8).

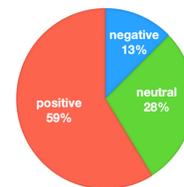
Figure 8 - Tip Sentiment breakdown

The final output table consists of 90,950 rows. A snippet of this table is shown below as Table 3. The total file size is 2.48MB and the total computation time using Hive is ~6 minutes (364.896 seconds).

Table 3. Output Snippet of Sentiment Analysis

| Country | State | Date | Sentiment | Count |
|---|---|---|---|---|
| USA | Nevada | 1/31/10 | negative | 5 |
| USA | Nevada | 1/31/10 | neutral | 17 |
| USA | Nevada | 1/31/10 | positive | 24 |
| USA | Nevada | 2/1/10 | negative | 3 |
| USA | Nevada | 2/1/10 | neutral | 2 |
| USA | Nevada | 2/1/10 | positive | 3 |
| USA | Nevada | 2/2/10 | positive | 5 |
| USA | Nevada | 2/2/10 | neutral | 1 |
| USA | Nevada | 2/3/10 | neutral | 3 |
| USA | Nevada | 2/3/10 | positive | 3 |
| USA | Nevada | 2/4/10 | positive | 2 |
| USA | Nevada | 2/5/10 | negative | 2 |
| USA | Nevada | 2/5/10 | neutral | 9 |
| USA | Nevada | 2/5/10 | positive | 7 |
| USA | Nevada | 2/6/10 | negative | 2 |
| USA | Nevada | 2/6/10 | neutral | 3 |
| … | … | … | … | … |

## 5.4 Yelp Star-Ratings
The third analysis is on "Yelp Star-Ratings". Yelp's platform allows its users to rate a business on a scale of 1-5 stars. The goal of this analysis is to identify trends among Yelp star ratings across the globe over time. The workflow to produce this analysis is illustrated in Figure 9 below.

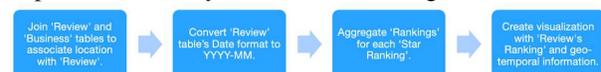
Figure 9. Workflow: Yelp Star-Ratings Analysis

This study was supported by Oracle Cloud Innovation Accelerator. Oracle Big Data Cloud Service was used in the data analysis.

We extracted data from the 'review' and 'business' primary tables. We joined together the 'review' and 'business' tables to associate geographic location with the 'review' table. The 'review' table's date format was then converted to YYYY-MM and the star-ratings (or rankings) were then aggregated. This geo-temporal analysis was visualized as an animation using Excel's 3D Maps to show how all of the businesses' star-ratings trend overtime across all geographic locations within our dataset (Figure 10).

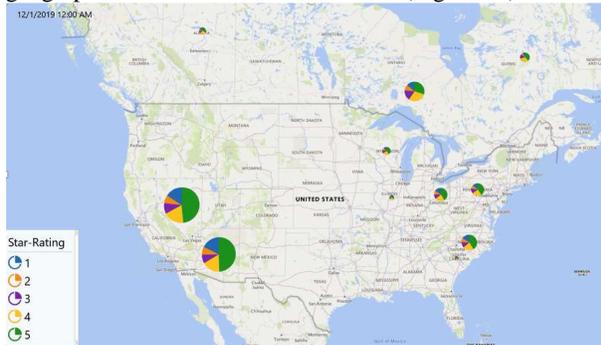

Figure 10 - Geo-Temporal Visualization of Yelp Star-Ratings

Key findings are:
i. Majority of businesses have 5-Star Ratings;
ii. Canadian businesses have more consistent differences between star-ratings;
iii. American businesses are more sporadic.

Figure 11 below compares total counts of star-ratings for Canada and the United States (Note the difference of scale).

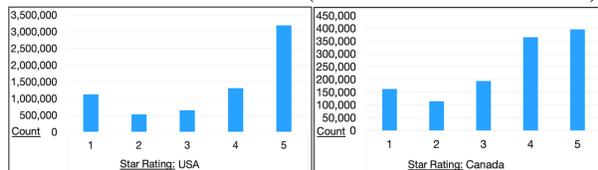

Figure 11 - Star-Ratings Comparison: USA vs Canada

The final output data for Yelp star-ratings consists of a total 9,832 rows. A snippet of this table is shown as Table 4 in the next column. The total file size is 1MB and the total computation time in Hive is ~1 minute (56.855 seconds).

Table 4. Output Snippet of Start Ratings Analysis

| Country | State | Date | Star-Rating | Count |
|---|---|---|---|---|
| USA | California | 6/1/10 | 1 | 1 |
| USA | California | 6/1/10 | 4 | 1 |
| USA | California | 6/1/10 | 5 | 4 |
| USA | California | 7/1/10 | 3 | 1 |
| USA | California | 7/1/10 | 4 | 1 |
| USA | California | 7/1/10 | 5 | 3 |
| USA | California | 9/1/10 | 5 | 1 |
| USA | California | 10/1/10 | 1 | 1 |
| USA | California | 11/1/10 | 5 | 1 |
| USA | California | 12/1/10 | 5 | 1 |
| USA | California | 1/1/11 | 5 | 3 |
| USA | California | 2/1/11 | 3 | 1 |
| … | … | … | … | … |

## 6. Conclusion

Ultimately, the data conversion and processing for this paper allowed for the use of different forms of data analysis. With the given analysis in the paper we are able to observe how different trends are occurring and just where these trends might lead to. The Yelp feature performance analysis allowed us to establish that reviews, users and tips have been declining since 2010. While check-ins were still being active up until the declining of that as well in the year 2015. The Yelp tip sentiment analysis presented us with information on how users have had more of a positive experience throughout both the USA and Canada. While the yelp star-ratings analysis provided us with an idea of just how different thinking Americans are when compared to Candians when rating businesses. Henceforth, with these data analyses we can conclude as well as establish that in the future yelp performance will and can continue to decline. Alternatively our analysis also shows that many people will continue to make use of the star ratings and leave good sentiment data to conclude just how well businesses are doing and will continue to do.

## References

[1] Ruchi Singh; Yashaswi Ananth; Jongwook Woo, "Big Data Analysis of Local Business and Reviews", in The 19th International Conference on Electronic Commerce: ICEC 2017, Pangyo, Seongnam, Korea, August 17- 18, 2017
[2] U. Griffo. "TwitterSentimentAnalysisAndN-gramWithHadoopAndHiveSQL.md." GitHub. https://gist.github.com/umbertogriffo/a512baaf63ce0797e175 (accessed Oct 26, 2020).
[3] Yelp, "Yelp Dataset." (Mar. 18th, 2020). Distributed by Yelp. https://www.yelp.com/dataset/ (accessed Oct. 8th, 2020).
[4] Ruchi Singh and Jongwook Woo, "Applications of Machine Learning Models on Yelp Data", Asia Pacific Journal of Information Systems (APJIS), VOL.29│NO.1│March 2019, pp35~49, ISSN 2288-5404 (Print) / ISSN 2288-6818 (Online)
[5] Kunal Pritwani, Knox Wasley, and Jongwook Woo, "Spark Big Data Analysis of World Development Indicators", in Global Journal of Computer Science and Technology (GJCST), Volume 18 Issue 3: 1-10, Version 1.0 Year 2018, ISSN 0975-4172

This study was supported by Oracle Cloud Innovation Accelerator. Oracle Big Data Cloud Service was used in the data analysis.